# Towards Human-AI Mutual Learning – A New Research Paradigm

Xiaomei Wang

Department of Industrial Engineering, University of Louisville, xiaomei.wang@louisville.edu

Xiaoyu Chen

Department of Industrial and Systems Engineering, University at Buffalo, xchen325@buffalo.edu

## 1 HUMAN-AI MUTUAL LEARNING

### 1.1 Definition and paradigm

Recent trends in human-artificial intelligence (AI) collaboration strive to achieve complementary expertise in human-AI teams [13,38], assuming that human and AI each hold unique expertise. It is expected that a successful human-AI team will have collaborative performance that exceeds the performance of humans or AI alone. To achieve this complementarity, we would highlight the concept of *mutual learning* between human and AI. The term *mutual learning* in the context of education usually describes learning through the sharing of knowledge and information between human individuals [17]. Correspondingly, we define human-AI mutual learning as the process where humans and AI agents preserve, exchange, and improve knowledge during human-AI collaboration.

Human-AI mutual learning is expected to open new directions for both researchers and practitioners in many domains, such as cognitive science, human-centered AI, reinforcement learning from human feedback (RLHF), etc. Figure 1 illustrates the opportunities of the human-AI mutual learning paradigm, which integrates both quantitative (blue) and qualitative (green) methodologies.

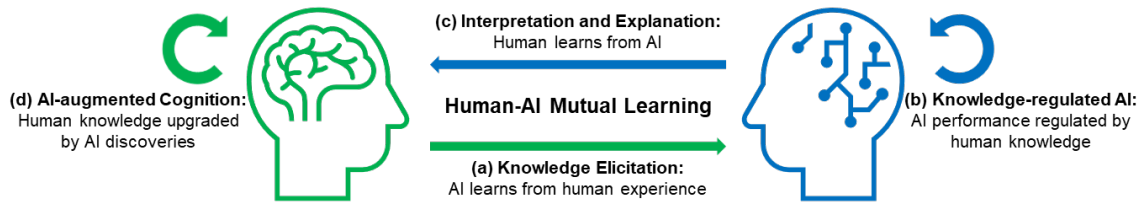

Figure 1: Human-AI mutual learning through knowledge elicitation and knowledge-regulated AI

### 1.2 Relevant Methodology

As shown in Figure 1(a), the human-AI mutual learning paradigm starts from knowledge elicitation from human experts. The field of cognitive psychology and human factors have made extensive contributions in developing and applying various knowledge elicitation techniques [7,33,36]. Some examples of widely used methods include hierarchical task analysis, cognitive work analysis, critical decision method, observation of representative tasks, protocol analysis, etc. [36].

The second step (Figure 1(b)) is to integrate human knowledge into AI model architecture and/or constraints to regulate the model training and testing process. For example, [4] proposes to quantitatively represent human decision rules (i.e., graph structured) into a partially connected neural network architecture to enforce AI by learning as an expert. On the other hand, studies incorporate domain knowledge in the form of physical laws to achieve knowledge-constrained AI, which has been explored in the fields of manufacturing, geoscience, material science, and several other domains [15,28]. Since existing work often uses physical laws, this approach is also referred to as "physics-constrained", "physics-informed", or "physics-aware". Recent research studies broadened the concept of knowledge constraints from physical laws to empirical and structural knowledge, e.g., safety constraints in safe reinforcement learning [8], RLHF [5], etc. More opportunities can be discovered by integrating unstructured human knowledge with AI models.

The rest steps aim to discover the benefit of AI to human knowledge. Such steps should be understood in two folds: to explain the black-box AI to elicit knowledge from AI (see Figure 1(c)); and to enable the effective and efficient representation of such elicited knowledge for human to comprehend (see Figure 1(d)). Many research studies have been reported in the past decades with the theme of explainable AI [23], which aims to explain how AI would make a decision (e.g., prediction, actions, etc.) under certain scenarios (i.e., local explanation [30]), or to summarize important patterns (e.g., features, trends, etc.) found by the AI model (i.e., global explanation [25]). These studies are crucial to enable the knowledge elicitation from AI models. However, it remains unclear how to better utilize the elicited knowledge from AI, which is often not natural to human knowledge structure. Hence, we also highlight the direction of AI-augmented cognition in Figure 1(d), which focuses on understanding how human cognition can be augmented by AI, in terms of decision making, situation awareness, mental workload, reasoning, sensemaking, problem solving, multitasking, etc. [20,29].

## 2 MOTIVATION AND RELEVANT WORK FROM DOMAIN EXAMPLES

Here we provide several examples from different domains to further explain the motivation of human-AI mutual learning.

### 2.1 Manufacturing

Domain knowledge has been playing an increasingly important role since the concept of manufacturing was established in ancient times. Psychology theories classify such knowledge into implicit and explicit knowledge [6], which play different but collaborative roles in decision-making processes. In modern manufacturing plants, the explicit knowledge is usually preserved in the format of standard operating procedures (SOP) (e.g., operation manuals, diagnosis and maintenance guidelines, etc.), product recipes (e.g., raw materials and their ratios, processes, and parameters, etc.), rule-based systems to support decision-making, etc. Many statistical and machine learning (ML) models have been developed in recent years which are governed by the aforementioned explicit knowledge [4,10,21,37]. However, representing implicit knowledge remains an open question, which prevents many knowledge-driven ML models from acting as human experts. As a direct result of the domain knowledge, learning from the behaviors of human experts during the decision-making process (e.g., diagnosing the root cause for a manufacturing failure) is vitally important for AI to infer experts' implicit knowledge.

### 2.2 Healthcare

Clinical decision support systems (CDSS), designed to support clinician decision making, are often classified as knowledge-based (or expert system) or non-knowledge-based (or data-driven, ML-based system) [34]. Early CDSS are knowledge-based systems mainly consisting of IF-TEHN rules that simulate human reasoning. With the advancement of AI and ML, non-knowledge-based systems have gained more attention in recent years.

Knowledge-based CDSS have advantages where rules are a natural format to express knowledge in a domain, and thus their outputs or recommendations are easier to understand by clinicians. However some barriers prevented knowledge-based CDSS from being widely adopted, e.g., usability issues, difficulty in updating the knowledge base, and our limited understanding of the human diagnostic reasoning process [2,27]. Non-knowledge-based systems, having the advantages of leveraging big data, and eliminating the need for collecting and representing human knowledge, have been criticized for issues including transparency, robustness, and trust [1,32].

The idea of combining the advantages of knowledge-based and non-knowledge-based CDSS is attractive, though we have not seen much progress from this perspective. Our proposed paradigm could shed light on future CDSS that at the same time, learn from data, learn from human expert knowledge, and outputs results that are more interpretable to human decision makers. One example in intraoperative hypotension management can be seen in [31] where an interpretable neural network model was constructed based on the knowledge structure of an anesthesiologist and then trained by real surgical data. While maintaining the superior performance of neural networks, human knowledge is optimized by learning from data thanks to the interpretable model structure.

## 3 BENEFITS

### 3.1 Benefits for AI

Both the methodology and application development of AI are expected to benefit from the mutual learning paradigm. From the perspective of AI methodologies, enabling AI to directly learn from human behaviors can help discover more effective model structures. One can find such benefits from many modern AI methods, e.g., the convolutional structure of neural networks was motivated by the human vision system [19], and the computational attention mechanism was to mimic human visual attention [14], etc. Compared with complex human behaviors, existing techniques are still in their infant stage, leaving much room for AI to evolve. In terms of benefit to applications, incorporating human explicit and implicit knowledge into AI training and inference can not only enhance the interpretability of AI [39], but also effectively reduce the costs to train the AI model. For example, physics-informed ML models can effectively reduce the sample size requirement [11].

### 3.2 Benefits for human

The benefit of this approach for human is the potential to better understand and utilize AI. With improved understanding and interpretation, AI could support the knowledge gain of novices, and also support appropriate trust.

One work by author XW [35] explored ML-based decision support systems in the context of graduate admission. The study investigated what information users find useful in helping them understand the system outputs. The information that was deemed most helpful was feature importance – a visualization that shows what variables were used to train the ML models and their importance. Users highlighted that this information helps them understand whether the system may generate recommendations in a similar or comparable pattern to themselves, and make judgments on whether to trust the system. With our proposed approach, we may see similar interpretability benefits and training benefits – human experts can more easily compare the system outputs with their existing knowledge, and human novices can not only follow/ignore system recommendations but also gain knowledge and insights.

## 4 POTENTIAL ISSUES AND FUTURE RESEARCH AGENDA

### 4.1 Is human-AI mutual learning necessary?

Integrating domain knowledge with AI is a fundamentally and practically challenging topic due to its cross-disciplinary nature and the insufficient theories to support such investigation. However, we would argue the necessity of such mutual learning mainly due to two reasons: (1) an AI method to directly learn from human behaviors will transform the current training paradigm, making AI training process no longer a specialty of data scientists, but enabling all AI users to train personalized AIs with their own behaviors. Hence the popularization of AI will be greatly accelerated. (2) Constraining AI to make trustworthy, safe, and fair decision-making support has been a challenging problem for existing modeling structures [22]. An AI regulated by human knowledge through directly learning from human behaviors is expected to tackle such challenges, since the AI is incubated by human users.

### 4.2 Knowledge elicitation and representation challenges

Knowledge elicitation is, and has always been challenging. Human knowledge is not all correct. Knowledge, especially empirical knowledge, has limited applicability. For domains like healthcare, best clinical practices and guidelines that represent our state-of-the-art knowledge of diseases are constantly being updated. Knowledge is also not always explicit; some seemingly intuitive expert judgments could be the results of tacit knowledge that underlies their reasoning [16].

Humans are probably better than AI at dealing with incomplete or even conflicted information/knowledge. [18] highlights that experts can make good decisions despite incomplete, incorrect, and contradictory information, and have rich mental models to support sensemaking. Whether AI is good at taking those incomplete, "locally" applicable knowledge, and validating its correctness is unknown.

Knowledge representation is challenging. Expert knowledge elicited may not be represented in a way that can be directly integrated into ML models. For example, the critical decision method is an interview method to understand expertise [26]. Typical data analysis is qualitative coding, e.g., thematic analysis [9], which will identify codes and themes from the interviews. Such qualitative results are good when reported as stand-alone findings on human expertise, but may be difficult to be "numericalized" or "programmed" to be ML model structure or constraints.

### 4.3 Developing the AI training and representation methodology behind knowledge-regulated ML is challenging

In the past few years, AI training scheme has gradually evolved from offline training to online training, and currently to online training with human feedback (e.g., RLHF). However, the state-of-the-art RLHF training scheme still requires one to define case-specific forms of human feedback (e.g., rating on Likert scale, pass/fail, etc.) [3], which is not compatible with human natural behaviors as their intents are usually implicit. How to continuously learn from human behaviors remains an open question. On the other hand, although, theoretically speaking, neural network enjoys the universal approximation theorem [12], the representation power of neural networks still strongly depends on the model architectures and the optimizers [24]. With implicit intents, large action space, and limited sensing modalities, even observing, representing, and understanding human behaviors can pose fundamental challenges to existing learning and representation theories of AI. A new AI training and representation framework is required to achieve mutual learning.

### 4.4 In-depth multidisciplinary collaboration

Our proposed paradigm requires in-depth multidisciplinary collaboration between experts in the knowledge domain, human factors/HCI, and AI. Building up a research team like this is likely challenging but rewarding.